\documentclass[fleqn,usenatbib]{mnras}

\usepackage{newtxtext,newtxmath}

\usepackage{amsmath}
\usepackage{bm}
\usepackage[T1]{fontenc}

\DeclareRobustCommand{\VAN}[3]{#2}
\let\VANthebibliography\thebibliography
\def\thebibliography{\DeclareRobustCommand{\VAN}[3]{##3}\VANthebibliography}

\defcitealias{1974Book..Whitham}{W74}
\defcitealias{1982SoPh...76..239E}{ER82}
\defcitealias{1986NASCP2449..347E}{ER86}
\defcitealias{2011ApJ...728..147C}{C11}
\defcitealias{2011SoPh..272..119F}{F11}
\defcitealias{2013ApJ...766...55K}{K13}
\defcitealias{2014ApJ...789...48O}{ORT14}

\usepackage{graphicx}	
\usepackage{amsmath}	
\usepackage{amssymb}	
\usepackage[switch]{lineno}	

\def\vec#1{\ensuremath{\bm{#1}}}
\newcommand\dispfrac[2]{\ensuremath{\displaystyle\frac{#1}{#2}}}

\newcommand{\Alf}{Alfv$\acute{\rm e}$n}

\newcommand{\rhoi}{\ensuremath{\rho_{\rm i}}}
\newcommand{\rhoe}{\ensuremath{\rho_{\rm e}}}

\newcommand{\va}{\ensuremath{v_{\rm A}}}
\newcommand{\vai}{\ensuremath{v_{\rm Ai}}}
\newcommand{\vae}{\ensuremath{v_{\rm Ae}}}

\newcommand{\vgr}{\ensuremath{v_{\rm gr}}}
\newcommand{\vph}{\ensuremath{v_{\rm ph}}}

\title[Kink Wave Trains in Slabs]{Impulsively Generated Kink Wave Trains in Solar Coronal Slabs}

\author[M. Guo et al.]{
Mingzhe Guo,$^{1}$\thanks{E-mail: m.guo@sdu.edu.cn}
Bo Li,$^{1}$
Tom Van Doorsselaere,$^{2}$
Mijie Shi$^{1}$
\\
$^{1}$Shandong Provincial Key Laboratory of Optical Astronomy and Solar-Terrestrial Environment, Institute of Space Sciences, Shandong University, \\
Weihai 264209, China\\
$^{2}$Centre for mathematical Plasma Astrophysics, Department of Mathematics, KU Leuven, 3001 Leuven, Belgium
}

\date{Accepted XXX. Received YYY; in original form ZZZ}

\pubyear{2022}

\begin{document}
\label{firstpage}
\pagerange{\pageref{firstpage}--\pageref{lastpage}}
\maketitle

\begin{abstract}
We numerically follow the response of density-enhanced slabs to impulsive, localized, transverse velocity perturbations by working in the framework of ideal magnetohydrodynamics (MHD).
 Both linear and nonlinear regimes are addressed. 
 Kink wave trains are seen to develop along the examined slabs, sharing the characteristics that more oscillatory patterns emerge with time and that the apparent wavelength increases with distance at a given instant. 
 Two features nonetheless arise due to nonlinearity, one being a density cavity close to the exciter and the other being the appearance of shocks both outside and inside the nominal slab. These features may be relevant for understanding the interaction between magnetic structures and such explosive events as coronal mass ejections. 
 Our numerical findings on kink wave trains in solar coronal slabs are discussed in connection with typical measurements of streamer waves. 
\end{abstract}




\section{Introduction}
A variety of wave events has been reported in different magnetic structures in the solar atmosphere \citep[see][for a recent review]{2020ARA&A..58..441N}.
By swaying the axis of waveguides, kink waves tend to be readily
   identifiable in imaging observations and therefore have been abundantly documented
   (e.g., \citealt[][]{1999ApJ...520..880A,1999Sci...285..862N,2012ApJ...759..144T, 2012ApJ...751L..27W, 2013A&A...560A.107A};
   also the most recent review by \citealt{2021SSRv..217...73N}).
Sausage waves are usually reported in the lower portions of the solar atmosphere \citep[e.g.,][and references therein]{2012NatCo...3.1315M,2021SoPh..296..184G}, 
with their identification also available in coronal loops \citep{2016ApJ...823L..16T}.
Meanwhile, 
rapid quasi-periodic pulsations (QPPs) with periods of seconds to a couple of tens of seconds are usually attributed to fast sausage modes \citep[see e.g.,][for recent reviews]{2020SSRv..216..136L,2021SSRv..217...66Z}.
Though not easy to measure,
Alfv\'en waves have also been reported in both spectroscopic \citep[e.g.,][]{2009Sci...323.1582J} and imaging observations \citep[e.g.,][]{2020A&A...633L...6K}.

Classical modelling of waves/oscillations in the solar atmosphere
   usually considers a cylindrical or a slab equilibrium.
Theoretical studies of collective waves in magnetic cylinders in the solar context
   date back to the 1970s and 1980s \citep[e.g.,][]{1970A&A.....9..159R,1975IGAFS..37....3Z,1979A&A....76...20W,1982SoPh...75....3S, 1983SoPh...88..179E},
   and have been gaining new momentum since
   the first imaging observations of kink motions in coronal loops by TRACE \citep{1999ApJ...520..880A,1999Sci...285..862N}.
As a consequence, collective waves have been extensively examined as a possible mechanism
   for heating the solar corona
   (e.g., \citealt{2017A&A...602A..74H, 2017A&A...604A.130K, 2019ApJ...870...55G, 2021ApJ...908..233S};
   see \citealt{2015RSPTA.37340261A}
       and \citealt{2020SSRv..216..140V} for recent reviews).
Likewise, they have been placed in the context of coronal seismology whereby
    coronal parameters can be indirectly inferred
    (e.g., \citealt{2001A&A...372L..53N, 2008A&A...491L...9V, 2016SoPh..291..877G,2020ApJ...894L..23M};
    see also the recent reviews by \citealt{2012RSPTA.370.3193D} 
    and \citealt{2020ARA&A..58..441N}).
Waves in magnetic slabs have also been extensively investigated \citep[e.g.,][]{1982SvAL....8..132Z,1982SoPh...76..239E,1993SoPh..144..101M,1993SoPh..145...65M,2015ApJ...801...23L,2021SoPh..296...95Y}.
Though simple, modelling considering a slab equilibrium is
    intuitively associated with some coronal measurements.
Two well-known typical observations are related.
One is the sunward propagating waves, 
    namely the "tadpoles" in open magnetic structures during a flare event \citep[][]{2005A&A...430L..65V}.
The other is the cyclic transverse displacements
    of streamer stalks, namely streamer waves,
    which are believed to be the largest transverse motions in the heliosphere \citep[][]{2010ApJ...714..644C, 2011ApJ...728..147C, 2011SoPh..272..119F,2013ApJ...766...55K, 2020ApJ...893...78D}.

Impulsively excited fast magnetoacoustic wave trains are relevant
    for interpreting a considerable number of 
    propagating disturbances observed in the corona.
For instance, the quasi-periodic fast propagating (QFP) disturbances measured in
    EUV  \citep[e.g.,][]{2011ApJ...736L..13L,2014SoPh..289.3233L}
can be intuitively understood as such.
These fast wave trains are observed in various wavelength bands in the corona \citep[][and references therein]{2001MNRAS.326..428W,2020SSRv..216..136L}.
They are examined analytically by \citet[][]{1984ApJ...279..857R} and \citet[][]{2015ApJ...806...56O},
 and numerically by e.g., \citet[][]{2004MNRAS.349..705N, 2017ApJ...836....1Y, 2021MNRAS.505.3505K}.
As demonstrated in e.g., \citet[][]{1984ApJ...279..857R},
the temporal signature of axisymmetric wave trains consists of  distinct phases when observed sufficiently far from the exciter.
These properties are attributed to the dispersive properties of fast sausage modes and are manifested in the wavelet spectrum as "crazy tadpoles" \citep[][]{2004MNRAS.349..705N} or "boomerangs" \citep[][]{2021MNRAS.505.3505K}.
In addition,
dispersive properties of impulsively generated fast kink wave trains have also been examined in coronal cylinders by e.g., \citet[][]{2015ApJ...806...56O,2017ApJ...836....1Y}.
Kink wave trains in coronal holes,
which are modelled as coronal funnels,
 are also investigated by \citet[][]{2014A&A...568A..20P}.

Until now,
most studies associated with fast wave trains either focus on axisymmetric waves and their seismological potentials
or discuss kink waves in magnetic cylinders.
Although studies of fast kink wave trains in a zero-$\beta$ slab have been conducted by e.g., \citet[][]{1993SoPh..144..101M,1993SoPh..145...65M,2021MNRAS.505.3505K},
an examination of the finite plasma $\beta$ effect is still necessary since it may introduce slow mode modulation,
as manifested in the discussion on sausage waves by \citet[][]{2017ApJ...847L..21P}.
In addition,
it seems too ideal to consider such interactions with magnetic structures to be linear,
given that the exciters of fast wave trains tend to be associated with such fierce solar activities
  as flares and coronal mass ejections (CMEs).
However,
the effect of nonlinearity has not been considered in studies associated with kink wave trains in slab configurations.

In this study,
we examine the reaction of coronal slabs to impulsive, localized, transverse velocity perturbations,
considering both a $\beta=0$ and a finite-$\beta$ effect.
Meanwhile,
we examine both linear and nonlinear regimes. 
Section~\ref{sec_0beta} describes the zero-$\beta$ computation and the numerical results.
Section~\ref{sec_Fbeta} presents linear and nonlinear analyses in finite-$\beta$ MHD.
Possible applications to streamer waves are discussed in Section~\ref{sec_sw}.
The conclusions are given in Section~\ref{sec_conclusion}.

\section{Linear Computation in Zero-$\beta$ MHD}
\label{sec_0beta}

In this section, we consider a coronal slab model under the cold plasma assumption ($\beta=0$) to obtain some basic understanding.
We adopt linear, pressureless, ideal MHD, 
assuming a static equilibrium ($\vec{v}_0=0$). 
Working in a Cartesian coordinate system $(x,y,z)$, 
we take the equilibrium magnetic field $\vec{B}_0$ 
to be uniform and $z$-directed. 
We further assume that the equilibrium density ($\rho_0$) depends only on $x$ and takes a step profile, 
\begin{eqnarray}
  \rho_0(x) 
= \left\{
       \begin{array}{ll}
          \rhoi,   &  |x| \le d, 			\\
          \rhoe,   &  |x| > d,
       \end{array} 
  \right.
\label{eq_rho_prof_gen}
\end{eqnarray}
   where $d$ is the slab half-width.
We define the \Alf\ speed $v_{\rm A}(x)$ as $B_0/\sqrt{\mu_0 \rho_0(x)}$ with $\mu_0$ being the magnetic permeability 
in free space.
By $\vai$ and $\vae$ we refer to the values of $\va$
   evaluated with $\rhoi$ and $\rhoe$, respectively
   ($\vae^2/\vai^2=\rhoi/\rhoe$).    
For in-plane propagation ($\partial/\partial y\equiv 0$),
   the governing equations are well known to reduce
   to \citep[e.g.,][]{2005A&A...441..371T,2018ApJ...855...53L},
\begin{eqnarray}
   \dispfrac{\partial^2 \xi}{\partial t^2}
=  v^2_{\rm A}(x)
   \left(\dispfrac{\partial^2}{\partial z^2}
        +\dispfrac{\partial^2}{\partial x^2}
   \right) \xi,
\label{eq_vx-t}
\end{eqnarray}
where 
$\xi (x,z;t)$
is the transverse displacement.
The following initial conditions (ICs) are adopted,
\begin{eqnarray}
&   \xi(x,z;t=0)
  =0,
	\label{eq_par vx ic}  \\
&   \dispfrac{\partial\xi}{\partial t}(x,z;t=0)
  =v_0  
    \exp\left(-\dispfrac{x^2}{2\sigma^2_x}\right)
    \exp\left(-\dispfrac{z^2}{2\sigma^2_z}\right),
    \label{eq_vx ic} 
\end{eqnarray}
   where $v_0$ represents the magnitude of the initial velocity perturbation,
   and $\sigma_x$ ($\sigma_z$) characterizes its spatial extent in the $x$- ($z$-) direction. 
We follow the evolution of an open system by adopting 
   a uniform grid to discretize a domain 
   that is large enough to make boundary conditions irrelevant. 
Initiated with the ICs, 
   Equation~\eqref{eq_vx-t} is advanced with a finite-difference scheme 
   second-order accurate in both time and space 
   \citep[see][for details]{2016ApJ...833...51Y,2022ApJ...928...33L}. 
We make sure that neither grid spacing nor time step affects our numerical results. 
The wave behavior is determined by the set 
$[\rhoi/\rhoe,\sigma_x/d,\sigma_z/d]$, which is fixed at
$[3,1/\sqrt{2},1/\sqrt{2}]$. 
The magnitude $v_0$ is chosen such that the maximum displacement along
   the slab axis is $\approx 2.6~d$.  

\begin{figure*}
\centering
\includegraphics[width=2\columnwidth]{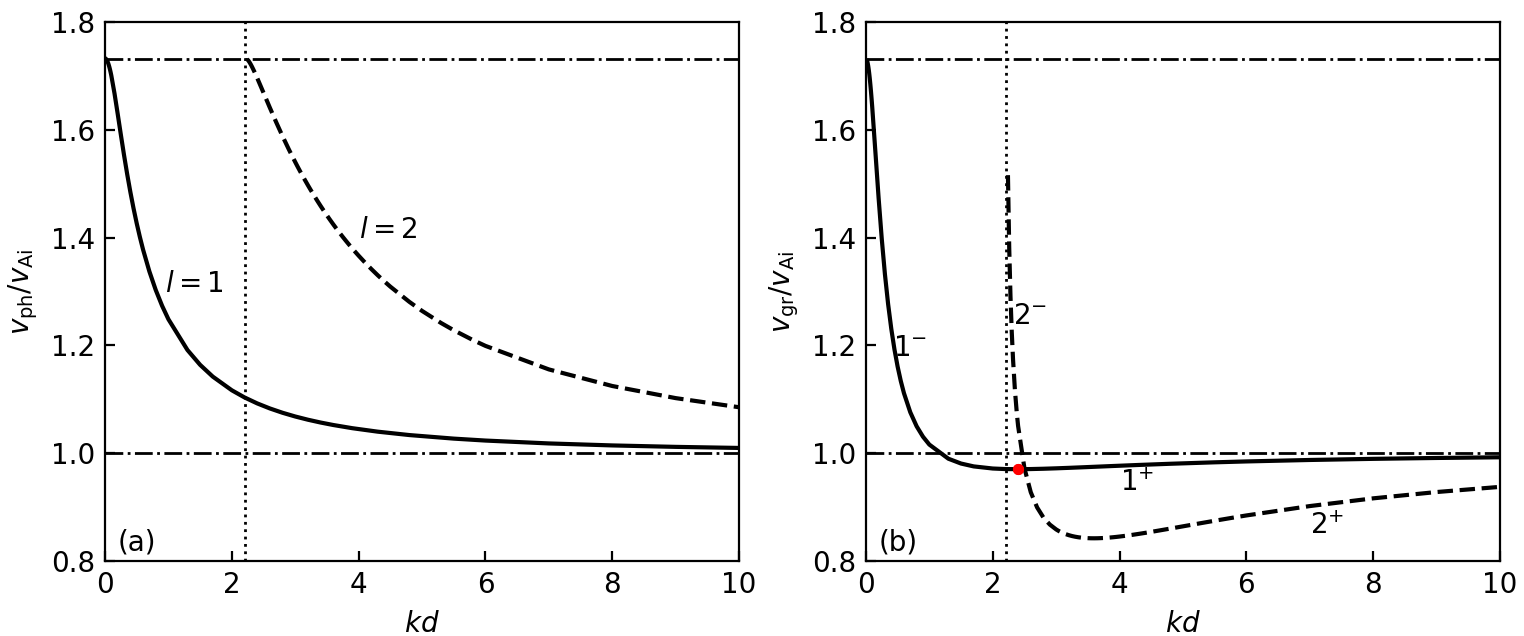}
\caption{
  Dependence on the dimensionless wavenumber ($kd$) of (a) the phase and (b) group
     speeds ($\vph$ and $\vgr$) of trapped (proper) kink eigenmodes 
     in a pressureless slab with a density contrast $\rhoi/\rhoe = 3$. 
  The transverse fundamental ($l=1$) and its first overtone ($l=2$)
     are shown by the solid and dashed curves, respectively.
  The horizontal dash-dotted lines represent the internal and external \Alf\ speeds
      ($\vai$ and $\vae$).
  Each curve in Figure~\ref{fig_vpvg}b is decomposed into two portions
      where $\vgr$ varies monotonically with $k$,
      with the red dot separating the relevant portions for $l=1$.
  }
\label{fig_vpvg}
\end{figure*}

\subsection{Dispersion Properties of Trapped Kink Modes}
\label{sec_sub_DR}

We start with an examination on the dispersion properties of kink modes
    by working with the well-known 
    dispersion relation 
    \citep[DR; e.g.,][]{1982SoPh...76..239E}
\begin{eqnarray}
   \sqrt{k^2-\dispfrac{\omega^2}{\vae^2}}
=  \sqrt{\dispfrac{\omega^2}{\vai^2}-k^2}
   \tan\left(
          d\sqrt{\dispfrac{\omega^2}{\vai^2}-k^2}
       \right),
\label{eq_DR}
\end{eqnarray}
    where the angular frequency $\omega$ and axial wavenumber $k$ characterize
    a Fourier component. 
Focusing on trapped modes ($\omega^2<k^2\vae^2$),
    we numerically solve Equation~\eqref{eq_DR} for given combinations
    of $[\rhoi/\rhoe,kd]$, thereby computing the axial phase 
    ($\vph\equiv\omega/k$) and group speeds
    ($\vgr\equiv d\omega/dk$).

Figure~\ref{fig_vpvg} presents, for $\rhoi/\rhoe=3$,
    the $k$-dependencies of 
    (a) $\vph$ and (b) $\vgr$.
Equation~\eqref{eq_DR} admits infinitely many branches of solutions,
    which we label with the transverse harmonic number $l =1,2,\cdots$. 
Figure~\ref{fig_vpvg} shows only 
    the transverse fundamental ($l=1$, the solid curves) 
    and its first overtone ($l=2$, dashed).
Examining the fundamental,
    one sees that trapped modes are allowed regardless of $k$.
In addition, both $\vph$ and $\vgr$
    start with $\vae$ at $kd\to0$
    and eventually approach $\vai$ when $kd\to\infty$.
However, $\vph$ does so by decreasing monotonically with $kd$, whereas
    $\vgr$ first decreases with $kd$ to some minimum
    (the red dot in Figure~\ref{fig_vpvg}b)
    before approaching $\vai$ from below. 
Moving on to the first overtone,
    one sees that trapped modes are permitted only when $k$ exceeds
    a cutoff axial wavenumber $k_{\rm cutoff,2}$.
For $k>k_{\rm cutoff,2}$, however, a monotonic (non-monotonic)
    $k$-dependence is seen for $\vph$ ($\vgr$) as happens 
    for the transverse fundamental. 
While only the $l=2$ branch is shown, all branches of trapped modes with $l\ge2$
    are subject to cutoffs ($k_{{\rm cutoff}, l}$), which themselves form
    a monotonically increasing sequence
    ($0=k_{{\rm cutoff},1} < k_{{\rm cutoff},2} < k_{{\rm cutoff},3}<\cdots$).    
In addition, the $\vgr-k$ curve for each $l$ can be divided into
    two portions, which will be labeled ``$l^-$'' (``$l^+$'')
    when $\vgr$ monotonically decreases (increases) with $k$ 
    (see Figure~\ref{fig_vpvg}b).

We place Figure~\ref{fig_vpvg} in the framework for understanding
    impulsively excited kink wave trains offered by 
    \citet[][]{2014ApJ...789...48O}.
While developed for a cylindrical equilibrium, 
    the \citet{2014ApJ...789...48O} framework applies here because
    the following mathematical properties are geometry-independent
        \footnote{
        The same physics applies also to sausage-type wave trains in both 
           cylindrical 
           (e.g., \citealt{2015ApJ...806...56O,2016ApJ...833...51Y}; also
           \citealt{1983Natur.305..688R})
           and Cartesian equilibria
           \citep[e.g.,][]{2004MNRAS.349..705N, 2013A&A...560A..97P,2021MNRAS.505.3505K}.}. 
The solutions to the eigenvalue problem pertinent to
    Equation~\eqref{eq_vx-t} on an open system are distinguished by the sign of 
    $k_{\rm e}^2\equiv k^2-\omega/\vae^2$, 
    with proper (improper) eigenmodes corresponding to 
    $k_{\rm e}^2<0$ ($k_{\rm e}^2>0$).
What we called ``trapped'' modes are therefore ``proper'' eigenmodes
    and will be referred to as such. 
The solution to our initial value problem
	can be formally expressed by the Fourier integral over $k$
	of individual components $\hat{\xi}(x,k;t)$.
The $z$-dependence of an initial perturbation is relevant 
    for determining $\hat{\xi}(x,k;t=0)$.
The $x$-dependence, on the other hand, determines how 
    $\hat{\xi}(x,k;t=0)$ is apportioned to proper and improper eigenmodes, 
    whose eigenfunctions depend on the equilibrium. 
An individual $\hat{\xi}(x,k;t)$ at a given $k$ therefore involves 
    two sets of contributions. 
The proper set consists of $L$ proper modes with $L$ being such that 
    $k_{{\rm cutoff},L}<k< k_{{\rm cutoff},L+1}$.
The improper set collects the contributions from improper modes with 
    $|\omega|$ extending continuously from $k \vae$ to infinity.

The signals sampled in the slab sufficiently far from the exciter involve
    only the proper contributions because improper modes 
    propagate laterally.
Consider the slab axis ($x=0$).
Our time-dependent results can be further placed in the framework detailed
    by Chapter~11 in \citet[][]{1974Book..Whitham},
    if we focus on those $[z,t]$ where $z$ and $t$ are large enough
    to ensure the applicability of
    the method of stationary phase (MSP). 
Seeing $[z,t]$ as given,
    the so-called ``stationary points'' in the MSP refer to
    the value(s) of $k$ that solves 
\begin{eqnarray}
 \vgr(k)=\dispfrac{z}{t}.
\label{eq_tmp1}
\end{eqnarray}   
If $[z,t]$ is seen as variable instead,
   then each solution to Equation~\eqref{eq_tmp1} defines
   a local axial wavenumber $k(z,t)$
   as a function of $z$ and $t$.
It then makes sense to define a local phase speed
   $\vph(z,t)$ with the DR by seeing $k$
   therein as $k(z,t)$.
By construction, the local group speed $\vgr(z,t)$
   is simply $z/t$.   
It further makes sense to define      
   a local frequency $\omega(z,t)=k(z,t)\vph(z,t)$
   and hence a local phase 
\begin{eqnarray}
\theta(z,t)=k(z,t) z-\omega(z,t) t. 
\label{eq_tmp2}
\end{eqnarray}
That the MSP applies at some $[z,t]$ 
   means that it applies to the corresponding local phase.
For brevity, we refer to such phases or space-time points as ``fully developed''.
We will visualize the notion that a phase needs to propagate for some time/distance
   to become fully developed, an intuitive expectation that has not been quantified
   to our knowledge. 
Nonetheless, the following notions apply to those fully developed
   phases or $[z,t]$.     
Firstly, the local values of $k(z,t)$ and $\omega(z,t)$
   are varying (fixed) as seen by an observer moving
   with the local phase (group) speed.
Secondly, a combination $[k(z,t),\omega(z,t)]$ defines
   one wavepacket, a concept that 
   \citet{1986NASCP2449..347E} introduced
   to the solar context.
Thirdly, for the problem at hand, Figure~\ref{fig_vpvg}
   indicates that a unique stationary point and hence a unique wavepacket
   can be determined from Equation~\eqref{eq_tmp1} at a given $[z,t]$
   only along a monotonic portion of the $\vgr(k)-k$ curve for an $l$.

\begin{figure*}
\centering
\includegraphics[width=2\columnwidth]{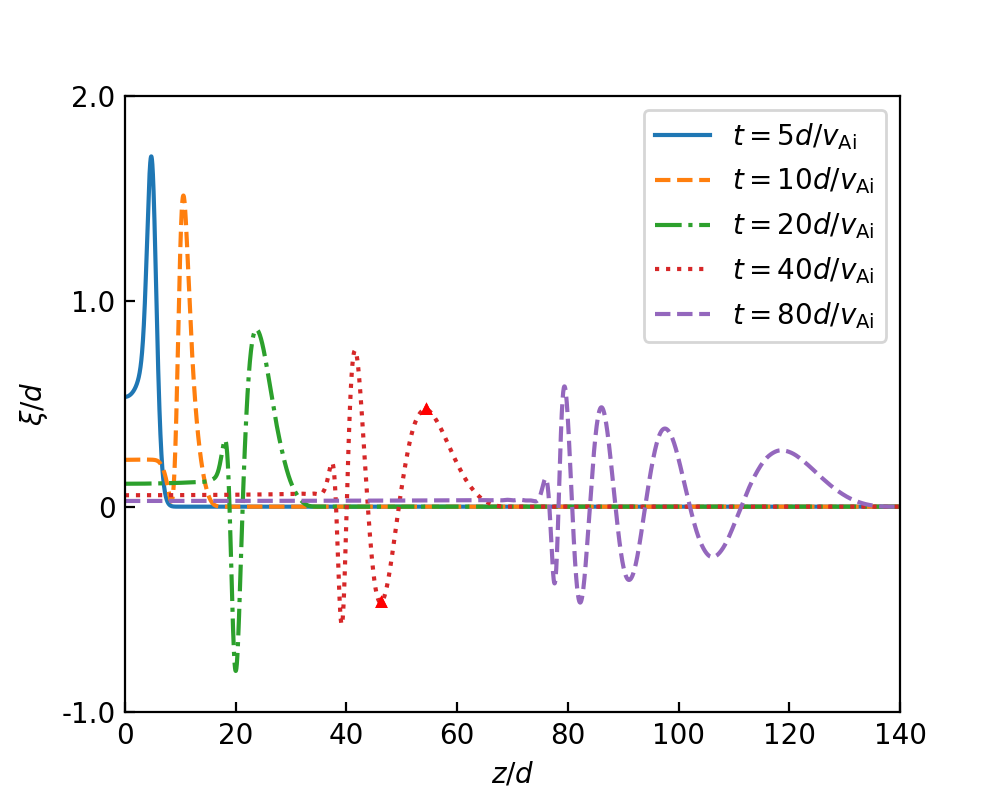}
\caption{
The $z$-distributions of the transverse displacement 
   $\xi(x,z,t)$ along the slab axis ($x=0$) at various instants as labelled. 
The red triangles 
   mark the outermost crest and trough at $t=40d/v_{\rm Ai}$.}
\label{fig_xi_z}
\end{figure*}

\subsection{Numerical Results}
\label{sec_sub_top-hat}

Figure~\ref{fig_xi_z} presents 
   the $z$-distributions of the transverse displacement ($\xi$) 
   along the slab axis ($x=0$) at various times as labeled.
Two properties can be observed from this figure.
One reads that more extrema appear as time proceeds.
The other is that
  the apparent wavelength tends to increase with distance in
  the snapshots of the transverse displacement.
This second signature is seismologically more useful since it cannot be taken for granted
   but arises when some constraints are placed on, say, 
   how the equilibrium is formulated
   and/or how localized the initial perturbations are.
Let crest~1 and trough~1 label the outermost crest and its trailing trough
   (see, e.g., the red triangles at $t=40~d/\vai$). 
With these two local phases as examples,
   we illustrate that fully developed phases are bound to 
   possess the second property if they involve only the wavepackets along 
   the $1^{-}$ portion in Figure~\ref{fig_vpvg}b. 
Provided this assertion, a unique wavepacket is guaranteed for a given phase 
   in a given instantaneous profile.   
Now that crest~1 leads trough~1, the instantaneous wavepacket at crest~1
   necessarily possesses a larger group speed.
One then deduces that the wavepacket at crest~1 necessarily possesses 
   a smaller local wavenumber and hence a longer local wavelength,
   given the $k$-dependence of $\vgr$. 
Turning this argument around, one further deduces that
   a sufficient condition for Figure~\ref{fig_xi_z} to reproduce the second property for 
   crest~1 and trough~1 is that these local phases are
   fully developed and the relevant wavepackets
   derive solely from the $1^{-}$ portion.

\begin{figure*}
\centering
 \includegraphics[width=2\columnwidth]{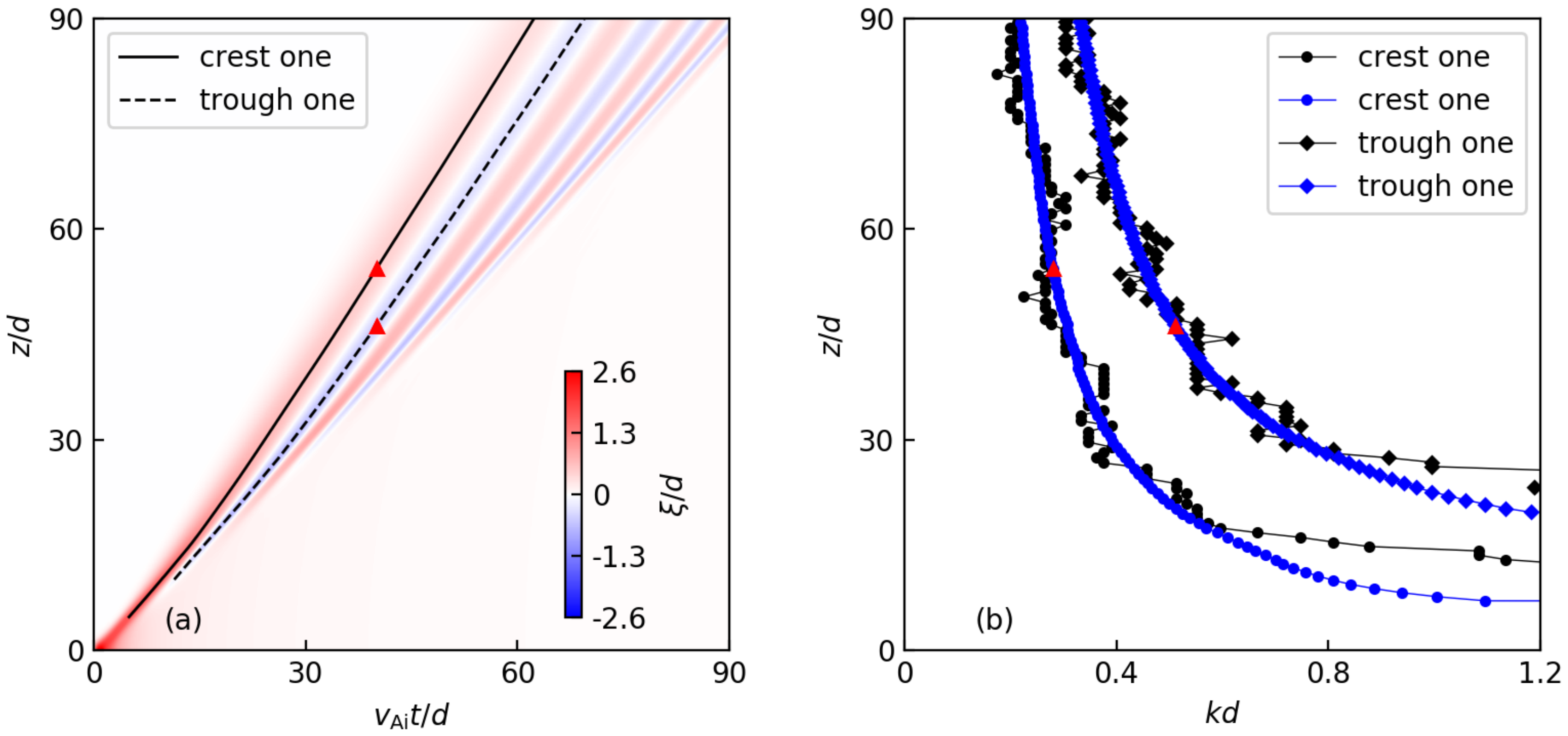}
 \caption{
 Left: Distribution in the $t-z$ plane of the transverse displacement ($\xi$)
       along the slab axis ($x=0$). 
       The outermost crest (trough) is highlighted with the solid (dashed) curve.
 Right: The local axial wavenumber $k$ as a function of $z$ for the outermost 
       crest and trough as determined with the local phase (the black symbols)
       and group (blue) speeds.
 The red triangles mark the outermost crest and trough at $t=40d/v_{\rm Ai}$.
 See text for details. 
}
\label{fig_xi_h_t}
\end{figure*}   

Two questions ensue.
One, how to quantify when/where a local phase is fully developed?
Two, how can we assert that only portion $1^{-}$ is relevant?
We will address question one shortly.
Accept for now that crest~1 and trough~1
   are fully developed in the curves at $t=40~d/\vai$ and $80~d/\vai$
   because they become so when propagating beyond $\sim20~d$ and $\sim30~d$, 
   respectively. 
Take the $t=40~d/\vai$ curve for instance.
One sees that the instantaneous group speeds ($z/t$) 
   at both phases
   exceed $\vai$, meaning that only portion~$2^{-}$
   may be potentially relevant in addition to $1^{-}$
   (see Figure~\ref{fig_vpvg}b).
Assuming that portion~$2^{-}$ is relevant, one expects that 
   crest~1 and/or trough~1 should be accompanied by a series of 
   short-scale ripples because the wavepackets 
   arriving at these locations necessarily possess large local
   axial wavenumbers if they are associated with portion~$2^{-}$.
This expectation, however, is not seen.    
     
We now address question one by showing 
   the $t-z$ distribution of 
   the transverse displacement $\xi(x=0,z;t)$ in Figure~\ref{fig_xi_h_t}a.
The emergence of crests and troughs is evident, and for simplicity
   we focus on the solid and dashed curves that thread
   crest~1 and trough~1, respectively.
These local phases at $t=40~d/\vai$ are further highlighted by
   the red triangles. 
We take crest~1 as example and demonstrate how to visualize 
   when/where it becomes fully developed by assuming
   that only portion $1^{-}$  is involved.
Briefly speaking, plotted in Figure~\ref{fig_xi_h_t}a
   is the numerical output $\xi(x=0,z_j;t_i)$
   for a uniform time sequence $\{t_i\}$
   and a uniform grid $\{z_j\}$.
The location $h_i$ and its temporal derivative
   $(dh/dt)_i$ can be found for crest~1 at an arbitrary $t_i$.
By construction, $(dh/dt)_i$ and $h_i/t_i$ represent
   the local phase and group speeds at $t_i$,
   and the values of $k_i$ as determined from 
       $\vph(k_{i, {\rm ph}})=(dh/dt)_i$
   and $\vgr(k_{i, {\rm gr}})=h_i/t_i$ should be the same
   if crest~1 is fully developed
   (see the discussions in Sect.\ref{sec_sub_DR}).
The blue and black circles in Figure~\ref{fig_xi_h_t}b
   then plot the $h_i$-dependencies of 
   $\{k_{i,{\rm ph}}\}$ and $\{k_{i,{\rm gr}}\}$, 
   respectively.
Repeating the same procedure further yields
   the blue and black diamonds that represent
   $\{k_{i,{\rm ph}}\}$ and $\{k_{i,{\rm gr}}\}$
   for trough~1. 
For both local phases, one sees that the black circles 
   somehow fluctuate, which we find difficult to quench.
Regardless, the circles in different colors
   experience some systematic deviation only below $\sim 20~d$, meaning
   that this is the critical distance for crest~1 to become fully developed. 
Likewise, one deduces a critical distance of $\sim30~d$ for trough~1, despite that it
   becomes discernible at a distance of $\sim10~d$ (see Figure~\ref{fig_xi_h_t}a). 
With this we illustrate that a given phase needs to propagate some time/distance
   to become fully developed.
More importantly, the circles (diamonds) in Figure~\ref{fig_xi_h_t}b 
   yield a consistent value of $k_i d\approx0.3$ ($\approx 0.55$)
   for crest ~1 (trough~1) at $t=40~d/\vai$
   (the red triangles in Figure~\ref{fig_xi_h_t}b).
This quantifies the increase of the apparent wavelength
   from trough~1 to crest~1 (see Figure~\ref{fig_xi_z}).
We remark that this quantification is not trivial, 
   because it does not make much sense to do this by
   counting the number of extrema per unit length.
  
\section{Linear and Nonlinear Computations in Finite-$\beta$ MHD}
\label{sec_Fbeta}
 
 \begin{table}
	\centering
	\caption{Parameters used in finite $\beta$ simulations. The internal slab density \rhoi~in ${\rm cm}^{-3}$, internal magnetic field $B_{\rm i}$ in G, temperature $T$ in Kelvin, and the half width $d$ in Mm.}
	\label{table}
	\begin{tabular}{ccccc} 
		\hline
	$\rho_{\rm i}$ & $B_{\rm i}$ & $T$ & $\rho_{\rm i}/\rho_{\rm e}$ & $d$\\
		\hline
	$3\times10^6$ & 1 & $1\times10^6$ & 3 & 100\\
		\hline
	\end{tabular}
\end{table} 

Now we proceed to the computations of a finite $\beta$ and continuous density distributions.
A more realistic equilibrium under the coronal condition is considered.
The initial parameters are listed in Table~\ref{table}. 
The plasma $\beta$ is 0.02 inside the slab at initial state.
The transverse density profile is chosen as

 \begin{eqnarray}
\rho(x)=\rho_{\rm e}+(\rho_{\rm i}-\rho_{\rm e})\zeta(x)~, 
\label{eq_rho_prof}
\end{eqnarray}
where 
\begin{eqnarray}
\zeta(x)=\dispfrac{1}{1+\left(x/d\right)^\alpha}~, 
\label{eq_rho_prof_zeta}
\end{eqnarray}
where $\alpha=10$ gives the steepness of the transverse density profile.
The initial density distribution has been presented in the leftmost column of Figure~\ref{fig_rho_contour}.
The temperature $T$ is uniform throughout the computational domain.
To maintain magnetostatic pressure balance, 
the magnetic field $\vec{B}=B(x)\hat{z}$ has a slight variation from the internal region of the slab to the external medium.

We employ the same initial perturbation as given by Equation~\eqref{eq_vx ic}.
In the following computations,
the magnitude of the initial velocity $v_0$ is chosen to be $0.1\vai$ and $3\vai$ in two different runs to distinguish between linear and non-linear regimes.

\subsection{Numerical Setup}
\label{sec_setup}   
We use the PLUTO code \citep{2007ApJS..170..228M} to solve the ideal MHD equations.
The HLLD Riemann solver is used to compute the numerical fluxes.  We use the piecewise parabolic method for spatial reconstruction and the second-order Runge-Kutta algorithm for time marching. 
The hyperbolic divergence cleaning method is adopted to maintain the divergence-free condition of the magnetic field. 
The two-dimensional computational domain is 
$[-100,100]d\times[-300,300]d$,
making sure that the boundaries are sufficiently far such that no reflections would influence the concerned region.
We consider a uniform mesh of 1000 points in the $x$-direction and 3000 uniformly spaced cells in the $z$-direction.
A convergence study with more grid cells shows no significant influence on the current results.

The boundary conditions are specified as follows. 
We fix the transverse velocities at both ends of the slab to be zero, 
while $v_z$, $B_x$ are set to have zero-gradients. 
Since the ends of the slab are sufficiently far from the initial perturbation,
the boundary set in the $z$-direction does not influence the concerned domain.
The other variables are fixed to the initial value. 
In addition,
we adopt outflow conditions to be the lateral boundaries 
in the $x$-direction.

\subsection{Numerical Results}
\label{sec_results_beta}   

 \begin{figure*}
\centering
 \includegraphics[width=2\columnwidth]{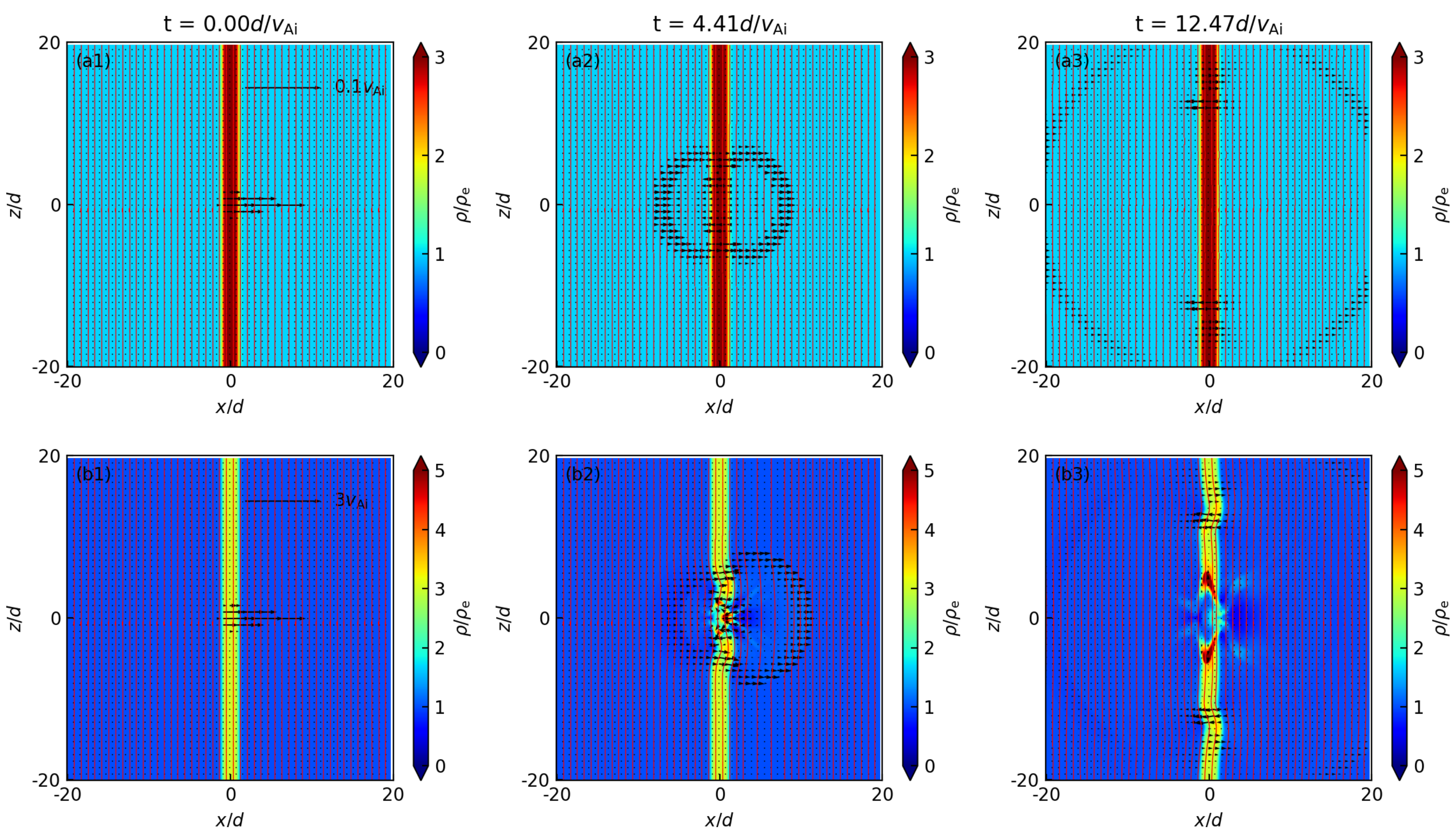}
 \caption{
 Evolution of normalized density distribution for linear (the upper row)
    and nonlinear (lower) computations at different instants as labeled.
 Magnetic lines of force (the red curves) 
    and velocity fields (the black arrows) are overplotted.
}
\label{fig_rho_contour}
\end{figure*}    

We conduct a preliminary examination of the dynamics of the magnetic slab with the help of density evolution maps.
Figure~\ref{fig_rho_contour} displays density snapshots for both
   linear (the upper row) and nonlinear (lower) 
   runs at the initial state and later instants.
Fast waves propagation can be observed in both runs,
   as shown by the velocity fields in Figure~\ref{fig_rho_contour}.
However,
no obvious density perturbations can be observed in the linear regime
   (the upper row of Figure~\ref{fig_rho_contour}).
In Figures~\ref{fig_rho_contour}b2 and \ref{fig_rho_contour}b3, however,
the density structure has noticeable displacements.
Kink waves characterized by transverse displacements of the slab axis can also be observed in Figure~\ref{fig_rho_contour}b2 and Figure~\ref{fig_rho_contour}b3.
Besides this,
one can find two other apparent signatures:
   the occurrence of a density cavity inside the slab near the exciter
       (clearly seen in Figure~\ref{fig_rho_contour}b3)
   and the propagation of density disturbances outside the slab.

\begin{figure*}
\centering
 \includegraphics[width=2\columnwidth]{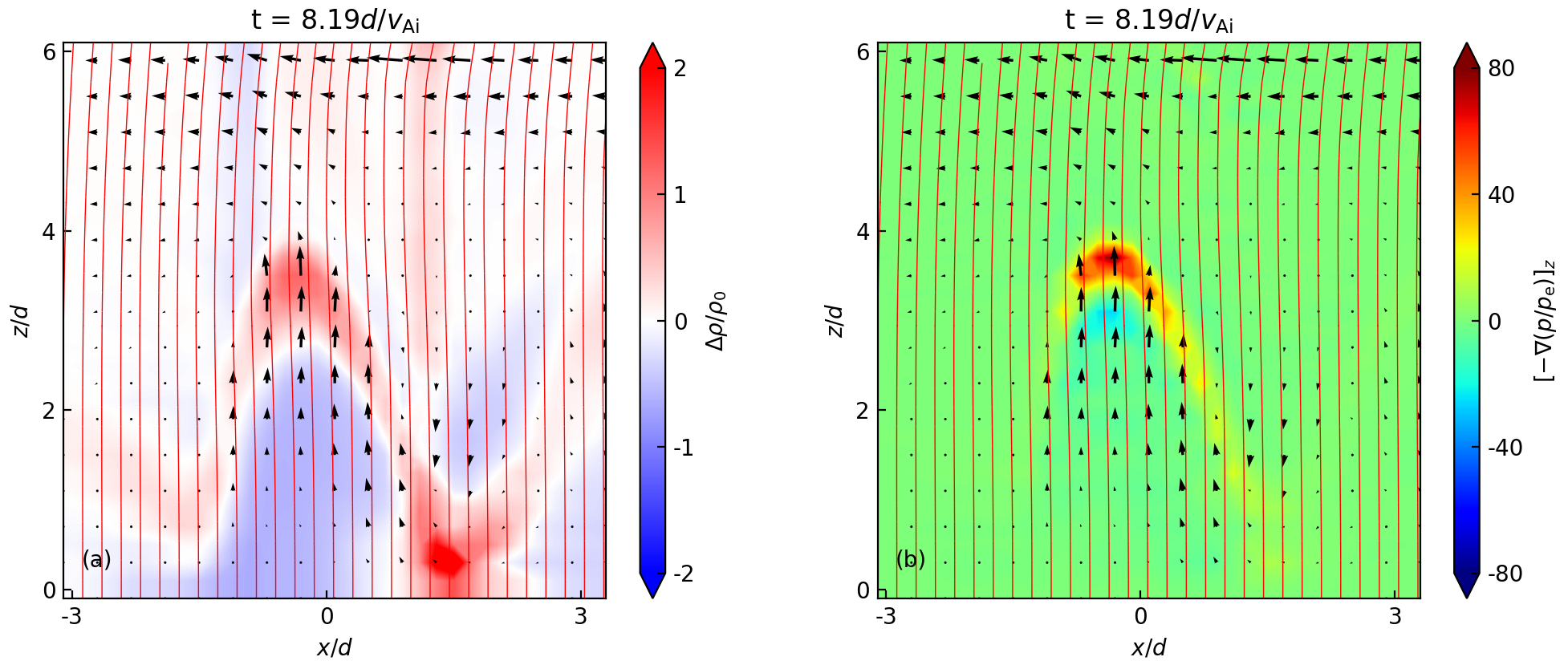}
 \caption{
 Left: Distribution of the density variation ($\Delta\rho=\rho(t)-\rho_0$) relative to the initial state $\rho_0$ at $t=8.19d/v_{\rm Ai}$.
 Right: Plots of the $z$-component of pressure gradients in the same region as in the left panel at $t=8.19d/v_{\rm Ai}$. The pressure is normalized by the external value $p_{\rm e}$. 
 Magnetic field lines (red) and velocity fields (black) are also overplotted in each panel.
}
\label{fig_rho_prs}
\end{figure*}

We first focus on the nonlinearity-induced signatures
   before proceeding to the analysis of the excited kink wave trains.
Figure~\ref{fig_rho_prs}a presents the density variations
   ($\Delta\rho=\rho(t)-\rho_0$) relative to the initial state at $t=8.19d/v_{\rm Ai}$.
Only the upper half of the concerned region is presented 
   in view of the relevant symmetry. 
One sees a density cavity characterized by some blue area inside
    the slab ($-d<x<d$) from $z=0$ to nearly $z=3d$.
Some density enhancement surrounding the cavity can also
    be seen.
The $z$-component of the pressure gradient defined by $\left[-\nabla(p/p_{\rm e})\right]_z$ has been plotted in Figure~\ref{fig_rho_prs}b.
One sees large pressure gradients accompanied by upward velocities in the dense area.
This means that the nonlinear initial pulse
    causes an increase in the local pressure gradient, 
    thereby inducing an upward flow and redistributing the local density.
A similar scenario has also been discussed by \citet[][]{2004ApJ...610..523T} in the reaction of a coronal loop to transverse perturbations.
In that case, a flow is also induced by the ponderomotive force, 
    causing the expansion of the loop.

\begin{figure*}
\centering
 \includegraphics[width=2\columnwidth]{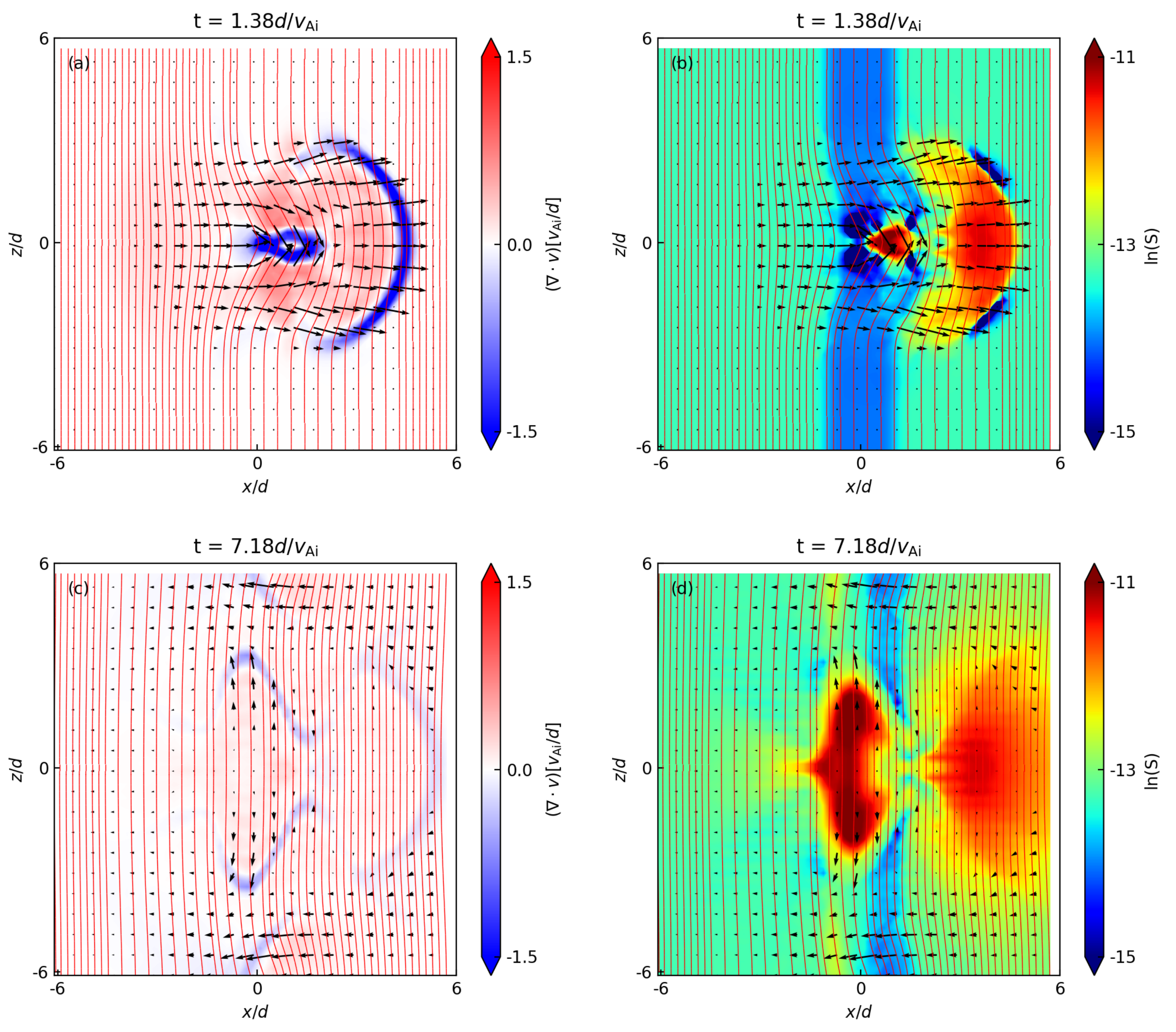}
 \caption{
 Snapshots of the velocity divergence ($\nabla\cdot \vec{v}$) and the entropy ($S=p\rho^{-\gamma}$) at different times as labelled. Magnetic field lines (red) and velocity fields (black) are also overplotted in each panel.
}
\label{fig_rho_shock_snap}
\end{figure*}

\begin{figure*}
\centering
 \includegraphics[width=2\columnwidth]{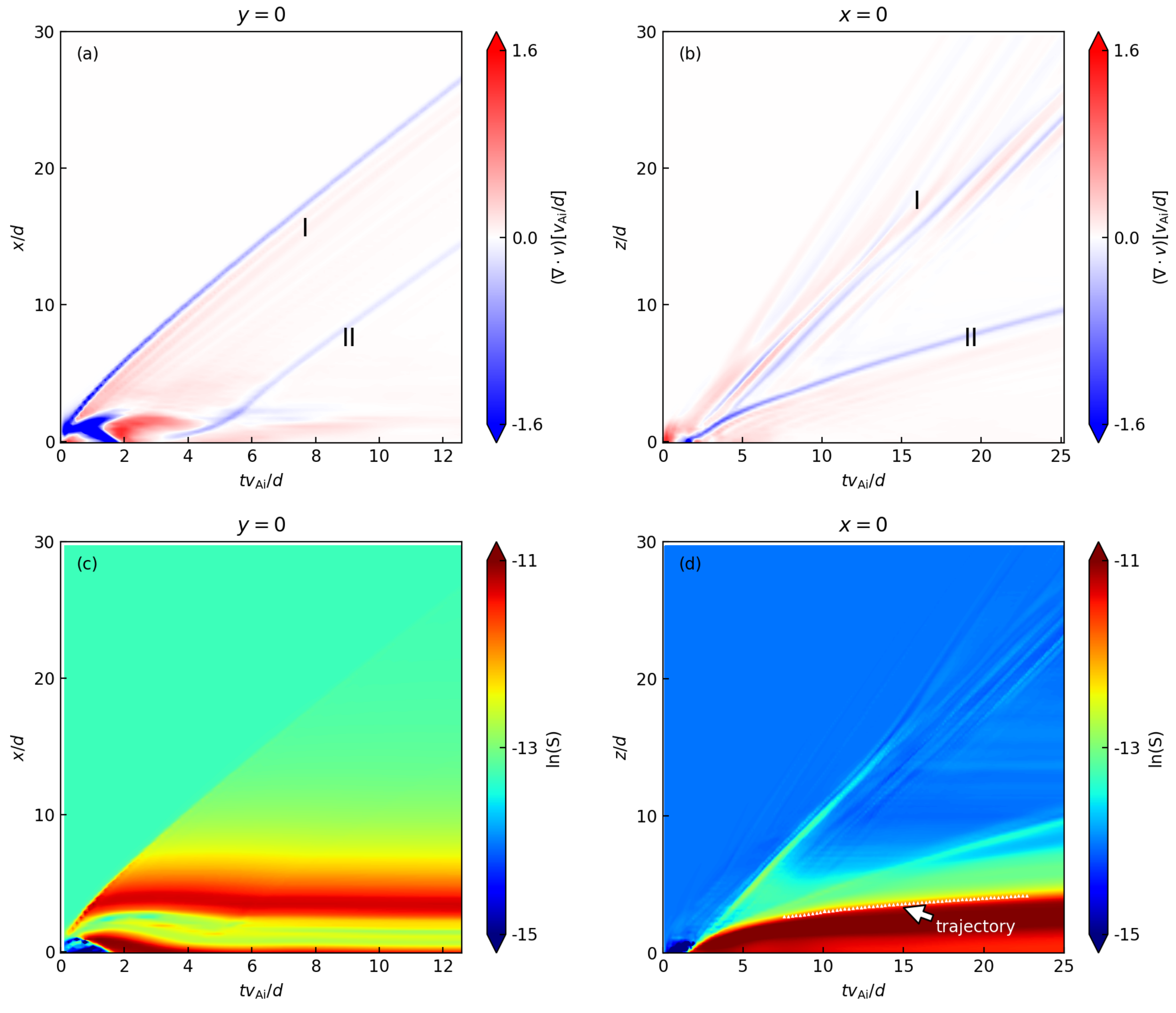}
 \caption{
 Upper: Time-distance map of the velocity divergence ($\nabla\cdot \vec{v}$) at (a) $y=0$ and (b) $x=0$. 
 Lower: Time-distance map of the entropy $S=p\rho^{-\gamma}$ at (c) $y=0$ and (d) $x=0$. Pressure $p$ and density $\rho$ are numerically dimensionless.
 White symbols in the lower panel represents the parcel trajectory of a given point $\left[tv_{\rm Ai}/d,z/d\right]=\left[10.08,3.01\right]$.
 See text for details.
}
\label{fig_rho_shock}
\end{figure*}

Another noteworthy signature in the nonlinear computation
    is the propagation of fast waves characterized by density enhancements.
As shown in Figure~\ref{fig_rho_contour}b2,
a wavefront indicated by the density enhancement moves outside the slab.
The wavefront is much clearer in the $x>0$ region since the initial perturbation is directed along the positive $x$-axis.
Figure~\ref{fig_rho_shock_snap} shows the distribution of
    the velocity divergence ($\nabla\cdot \vec{v}$) and the logarithmic entropy ($\ln S$) at two different instants as labeled. 
Here a variable $S=p\rho^{-\gamma}$ is introduced to measure the specific entropy
    with $\gamma=5/3$ being the adiabatic index.
One sees that a wavefront disturbs the magnetic field lines in Figure~\ref{fig_rho_shock_snap}a,b,
introducing discontinuous variations in velocity divergence and entropy.
This shows the presence of a shock. 
The downstream magnetic field is refracted away from the shock normal,
showing a fast shock property.
At a later time,
one can also find a wavefront inside the slab in Figure~\ref{fig_rho_shock_snap}c,d.
The velocity is directed almost along the magnetic field lines,
and the downstream magnetic field is refracted towards the wavefront normal,
showing a possible sign of a slow shock.
However,
this signature is mixed with the upflow shown in Figure~\ref{fig_rho_prs}b.
To recognize the slow shock,
it is needed to trace their evolution at given positions.
Figure~\ref{fig_rho_shock}a,c present the temporal evolution of velocity divergence and entropy at $x=0$ and $y=0$.
One sees the fast shock front characterized by a negative
    velocity divergence (blue stripes) in the region I of Figure~\ref{fig_rho_shock}a,
    followed by rarefactions characterized by a bundle of red stripes.
The discontinuity of the entropy in Figure~\ref{fig_rho_shock}c proves again that the outside propagating disturbances steepen into shocks.
The propagation speed of the shock front measured from Figure~\ref{fig_rho_shock}c is larger than the local fast speed. 
This confirms again a fast shock.
Region II in Figure~\ref{fig_rho_shock} reveals a fast wave reflected by the slab lateral boundaries.
In addition,
one can now find the slow shock hidden by the pressure-induced flow inside the slab.
A similar distribution of red and blue stripes can also be found in region II of Figure~\ref{fig_rho_shock}b,
manifesting a shock wave moving along the slab.
Due to the pressure-induced flow,
however,
the shock front can only be recognized after about $t=17d/v_{\rm Ai}$ in Figure~\ref{fig_rho_shock}d.
Again,
the propagation speed of the shock front measured from Figure~\ref{fig_rho_shock}d is larger than the local sound speed but much smaller than the local \Alf~speed. 
This confirms a slow shock.
Seen from Figure~\ref{fig_rho_shock}c,
the density cavity shows no oscillatory property,
but shrink to $x=0$ after $\sim 4v_{\rm Ai}/d$.
Note that the enhancements of entropy in about $z<4d$ in Figure~\ref{fig_rho_shock}d are induced by the density cavity.
The trajectory of a given parcel
    ($\left[tv_{\rm Ai}/d,z/d\right]=\left[10.08,3.01\right]$) is traced,
    shown by the white symbols in Figure~\ref{fig_rho_shock}d \footnote{This parcel trajectory is the projection of the trajectory in the $\left[x,z\right]$ space on the $z$-direction.
    It is very close to the real one since the $v_x$ is nearly zero in the time interval we considered.}. 
The parcel trajectory coincides with the boundary of the density cavity,
     meaning that the cavity is associated with plasma flows rather than being a genuine wave.
The red and blue stripes in the region I in Figure~\ref{fig_rho_shock}b are a signal of fast wave trains in the slab,
manifesting the compressibility of kink waves.
Similar slow shocks can also be observed when perturbing a slab with symmetric initial perturbations,
as discussed by \citet[][]{2017ApJ...847L..21P}.

\begin{figure*}
\centering
 \includegraphics[width=2\columnwidth]{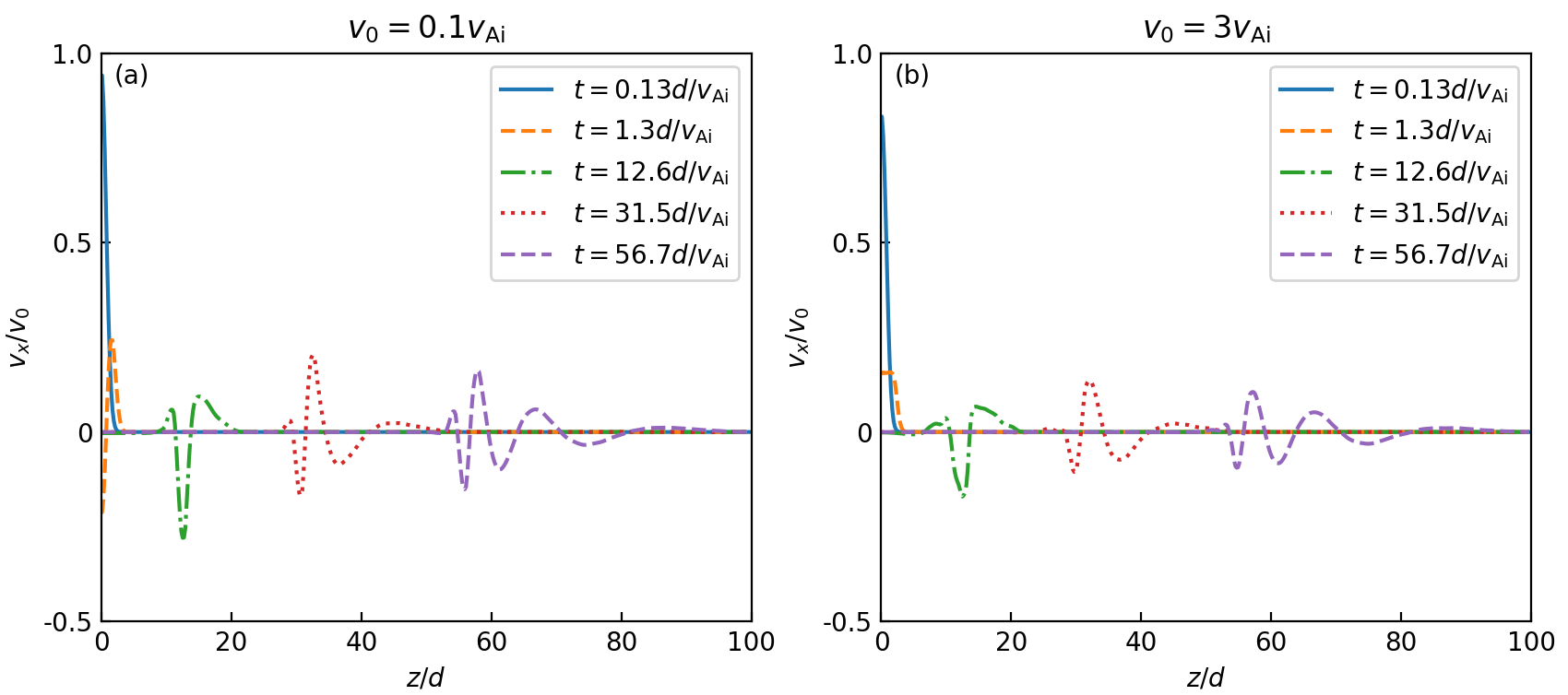}
 \caption{
 The $z$-distributions of the normalized transverse velocity 
   $v_x/v_0$ along the slab axis ($x=0$) at various instants as labelled. 
   The time $t$ is measured in units of $d/v_{\rm Ai}$.
}
\label{fig_xi_z_beta}
\end{figure*}

Now we examine the properties of the excited kink wave trains in the slab with a finite plasma $\beta$. 
Similar to Figure~\ref{fig_xi_z},
we present transverse velocity $v_x$ for both linear and nonlinear regimes at different times in Figure~\ref{fig_xi_z_beta}
\footnote{ One may find that the profiles in Figure~\ref{fig_xi_z_beta} differ from those in Figure~\ref{fig_xi_z},
especially in the initial stage. The reason is that different instants are chosen in these two figures.}.
The transverse velocity is normalized by the initial velocity amplitude $v_0$ for both cases.
One sees that the wave properties of the two regimes are similar at given instants,
    showing no significant influence of the amplitude of initial perturbations on the wave profiles,
    despite the density variations and shocks in the nonlinear computation.
Meanwhile,
the properties that more oscillatory patterns emerge with time and wavelength increases at a given time show up as happens 
in the $\beta=0$ calculation in Figure~\ref{fig_xi_z}.
The understanding of these properties has been illustrated in Section~\ref{sec_0beta}.
Note that the wave profiles show no steepening as time evolves.
The reasons are twofold.
First, the wave energy distributes to larger areas as time proceeds,
    leading to a decrease in the velocity amplitude.
Second,
as demonstrated in previous sections,
the head of the wave train moves faster than its tail,
thus no steepening can be observed.
We note that the continuous density distribution and the occurrence of fast/slow shocks do not influence the oscillatory properties of fast kink wave trains in the slab.

\section{Potential Application to Streamer Waves}
\label{sec_sw} 

Kink wave trains in slabs are intuitively related to streamer wave observations.
Actually,
three signatures have been reported in imaging observations of streamer waves \citep[e.g.,][]{2010ApJ...714..644C,2013ApJ...766...55K}. 
Signature 1, more oscillatory patterns emerge as time progresses. 
Signature 2, the apparent wavelength increases with distance at a given time. 
Both signatures were evident in the sequence of LASCO C2 and C3 running difference images \citep[][Figure~4]{2010ApJ...714..644C}.
Signature 3, the cyclic motion tends to be rapidly attenuated at small heliocentric distances, 
evidenced in the STEREO/COR1 measurements 
(see \citeauthor{2013ApJ...766...55K} \citeyear{2013ApJ...766...55K}; Figures~5 and 6).
Our numerical results have shown that Signatures~1 and 2 are compatible with a simple picture where streamer waves are impulsively excited kink wave trains in density-enhanced stalks even when the propagation is strictly two-dimensional and when the equilibrium is axially uniform,
regardless of the reality of a finite plasma $\beta$ and nonlinear effects.

Our nonlinear calculation seems not appropriate to reproduce the observational properties of Signature 3 close to the exciter.
Due to the occurence of the density cavity,
it would be not easy to recognize a magnetic slab structure there,
thus Signature 3 cannot be discussed.
However,
the observational events associated with Signature 3,
such as the streamer waves reported by \citet{2013ApJ...766...55K},
are excited by external pulses (e.g., CMEs) at streamer helmet bases.
Therefore the density cavity can not be observed in the streamer stalks.
However,
if a CME impinges on a streamer stalk from the side,
the appearance of a density cavity would be possible.
In this case,
however,
detections of such a density cavity would still be a challenge.
Firstly,
the intensity of the density cavity is determined by the magnitude of perturbations and interaction time between CMEs and streamer stalks.
Therefore high spatial and temporal resolutions of instruments are necessary.
Secondly,
the intensity distribution of streamer stalks tends to be diffuse, 
and the density contrast between streamer stalks and background corona is not large enough (see Figure~2 in \citeauthor{2011ApJ...728..147C}~\citeyear{2011ApJ...728..147C}).
This observational fact may also introduce difficulties in recognizing a density cavity in a streamer stalk.
In addition,
CMEs probably do not impinge on the stalks perpendicularly,
and the potential flow may fall back due to gravity.
These possibilities also influence convincing detections of a density cavity in streamer stalks.

\section{Conclusions}
\label{sec_conclusion}

In this paper we examined the reaction of coronal slabs to both linear and nonlinear localized transverse velocity perturbations, considering both plasma $\beta=0$ and a finite plasma $\beta$ effect.
All computations confirm the presence of fast kink wave trains in the slab. 
The wave trains present the emergence of more oscillatory patterns with time and the apparent increase of wavelength with distance at a given instant, 
regardless of the nonlinearity and the reality of a finite plasma $\beta$.
Particularly,
a density cavity and enhancement near the exciter are observed in the nonlinear computation.
In addition,
fast/slow shocks are also obtained in the nonlinear regime.

Potential applications of the current results of kink wave trains in coronal slabs are related to the typical observations of streamer waves. 
All the computations can reproduce the observational signatures of streamer waves that more oscillatory patterns emerge as time marches and the apparent wavelength increases with distance at a given time.
Our linear calculations can also reproduce the rapid damping of streamer waves at small heliocentric distances,
regardless of the reality of a finite plasma $\beta$ and density distributions.
The nonlinear calculations with a finite $\beta$ consideration predict the occurrence of a density cavity in streamer stalks
when the fierce activities, such as CMEs interact directly with streamer stalks rather than the streamer helmet bases.

\section*{Acknowledgements}

We gratefully acknowledge ISSI-BJ for supporting the international team "Magnetohydrodynamic wavetrains as a tool for probing the solar corona".
This work was supported by the National Natural Science Foundation of China (41974200, 11761141002, 41904150) and the European Research Council (ERC) under the European Union's Horizon 2020 research and innovation programme (grant agreement No.724326).

\section*{Data Availability}

The data underlying this article are available in the article and in its references.



\bibliographystyle{mnras}
\bibliography{kink_slab_streamer_waves} 








\bsp	
\label{lastpage}
\end{document}